\magnification \magstep1
\raggedbottom
\openup 4\jot
\voffset6truemm
\pageno=1
\headline={\ifnum\pageno=1\hfill\else
\hfill {\it Non-local boundary conditions for massless
spin-${1\over 2}$ fields} \hfill \fi}
\rightline {DSF preprint 96/36}
\centerline {\bf NON-LOCAL BOUNDARY CONDITIONS FOR}
\centerline {\bf MASSLESS SPIN-${1\over 2}$ FIELDS}
\vskip 1cm
\leftline {Giampiero Esposito}
\vskip 1cm
\noindent
Istituto Nazionale di Fisica Nucleare, Sezione di Napoli,
Mostra d'Oltremare Padiglione 20, 80125 Napoli, Italy;
\vskip 0.3cm
\noindent
Dipartimento di Scienze Fisiche, Mostra d'Oltremare 
Padiglione 19, 80125 Napoli, Italy.
\vskip 1cm
\noindent
{\bf Abstract.} This paper studies the 1-loop 
approximation for a massless spin-1/2
field on a flat four-dimensional Euclidean background 
bounded by two concentric 3-spheres, when non-local boundary
conditions of the spectral type are imposed. The use of
$\zeta$-function regularization shows that the conformal
anomaly vanishes, as in the case when the same field
is subject to local boundary conditions involving projectors.
A similar analysis of non-local boundary conditions can be
performed for massless supergravity models on manifolds
with boundary, to study their 1-loop properties.
\vskip 100cm
\leftline {\bf 1. Introduction}
\vskip 0.3cm
\noindent
The quantum theory of fermionic fields can be expressed,
following the ideas of Feynman, in terms of amplitudes
of going from suitable fermionic data on a spacelike surface
${\cal S}_{I}$, say, to fermionic data on a spacelike
surface ${\cal S}_{F}$. To make sure that the quantum 
boundary-value problem is well posed, one has actually to
consider the Euclidean formulation, where the boundary
3-surfaces, $\Sigma_{I}$ and $\Sigma_{F}$, say, may be
regarded as (compact) Riemannian 3-manifolds bounding a
Riemannian 4-manifold. In the case of massless spin-1/2
fields, which are the object of our investigation, one thus
deals with transition amplitudes
$$
{\cal A}[{\rm boundary \; data}]=\int {\rm e}^{-I_{E}}
{\cal D}\psi \; {\cal D}{\widetilde \psi}
\eqno (1.1)
$$
where $I_{E}$ is the Euclidean action functional, and
the integration is over all massless spin-1/2 fields 
matching the boundary data on $\Sigma_{I}$ and $\Sigma_{F}$. 
The path-integral representation of the quantum
amplitude (1.1) is then obtained with the help of Berezin
integration rules, and one has a choice of non-local [1] or
local [2] boundary conditions. The mathematical foundations
of the former lie in the theory of spectral asymmetry and
Riemannian geometry [3], and their formulation can be described
as follows. In two-component spinor notation, a massless
spin-1/2 field in a 4-manifold with positive-definite 
metric is represented by a pair
$\Bigr(\psi^{A},{\widetilde \psi}_{A'}\Bigr)$ of independent
spinor fields, not related by any spinor conjugation. Suppose
now that $\psi^{A}$ and ${\widetilde \psi}^{A'}$ are expanded
on a family of concentric 3-spheres as
$$
\psi^{A}={1\over 2\pi}\tau^{-{3\over 2}}\sum_{n=0}^{\infty}
\sum_{p,q=1}^{(n+1)(n+2)}\alpha_{n}^{pq}
\Bigr[m_{np}(\tau)\rho^{nqA}+{\widetilde r}_{np}(\tau)
{\overline \sigma}^{nqA}\Bigr]
\eqno (1.2)
$$
$$
{\widetilde \psi}^{A'}={1\over 2\pi}\tau^{-{3\over 2}}
\sum_{n=0}^{\infty}\sum_{p,q=1}^{(n+1)(n+2)}
\alpha_{n}^{pq}\Bigr[{\widetilde m}_{np}(\tau)
{\overline \rho}^{nqA'}+r_{np}(\tau)\sigma^{nqA'}\Bigr].
\eqno (1.3)
$$
With a standard notation,
$\tau$ is the Euclidean-time coordinate which
plays the role of a radial coordinate, and the block-diagonal
matrices $\alpha_{n}^{pq}$ and the $\rho$- and 
$\sigma$-harmonics are described in detail in [4].
One can now check that the harmonics $\rho^{nqA}$ 
have positive eigenvalues for the intrinsic
three-dimensional Dirac operator on $S^{3}$:
$$
{\cal D}_{AB} \equiv {_{e}n_{AB'}} \; e_{B}^{\; \; B'j}
\; { }^{(3)}D_{j} 
\eqno (1.4)
$$ 
and similarly for the harmonics $\sigma^{nqA'}$ and the
Dirac operator
$$
{\cal D}_{A'B'} \equiv {_{e}n_{BA'}} \; e_{B'}^{\; \; \; Bj}
\; { }^{(3)}D_{j} .
\eqno (1.5)
$$ 
With our notation, ${_{e}n_{AB'}}$ is the Euclidean normal
to the boundary, $e_{B}^{\; \; B'j}$ are the spatial 
components of the two-spinor version of the tetrad, and
${ }^{(3)}D_{j}$ denotes three-dimensional covariant
differentiation on $S^{3}$ [1, 2, 4].
By contrast, the harmonics ${\overline \sigma}^{nqA}$ and
${\overline \rho}^{nqA'}$ have negative eigenvalues for
the operators (1.4) and (1.5) respectively.

The so-called {\it spectral} boundary conditions
rely therefore on a non-local operation, i.e. the separation of
the spectrum of a first-order elliptic operator 
(our (1.4) and (1.5))
into a positive and a negative part. They require that half of
the spin-1/2 field should vanish on $\Sigma_{F}$, where this half
is given by those modes $m_{np}(\tau)$ and $r_{np}(\tau)$ which
multiply harmonics having positive eigenvalues 
for (1.4) and (1.5) respectively.
The remaining half of the field should vanish on $\Sigma_{I}$, and
is given by those modes ${\widetilde r}_{np}(\tau)$ and
${\widetilde m}_{np}(\tau)$ which multiply harmonics having
negative eigenvalues for 
(1.4) and (1.5) respectively. One thus writes [4]
$$
\Bigr[\psi^{A}_{(+)}\Bigr]_{\Sigma_{F}}=0
\Longrightarrow \Bigr[m_{np}\Bigr]_{\Sigma_{F}}=0
\eqno (1.6)
$$
$$
\Bigr[{\widetilde \psi}^{A'}_{(+)}\Bigr]_{\Sigma_{F}}=0
\Longrightarrow \Bigr[r_{np}\Bigr]_{\Sigma_{F}}=0
\eqno (1.7)
$$
and
$$
\Bigr[\psi^{A}_{(-)}\Bigr]_{\Sigma_{I}}=0 
\Longrightarrow 
\Bigr[{\widetilde r}_{np}\Bigr]_{\Sigma_{I}}=0
\eqno (1.8)
$$
$$
\Bigr[{\widetilde \psi}^{A'}_{(-)}\Bigr]_{\Sigma_{I}}=0
\Longrightarrow
\Bigr[{\widetilde m}_{np}\Bigr]_{\Sigma_{I}}=0 .
\eqno (1.9)
$$
Massless spin-1/2 fields are here studied since they provide
an interesting example of conformally invariant field
theory for which the spectral boundary conditions (1.6)--(1.9)
occur naturally already at the classical level [3].

Section 2 is devoted to the evaluation of the
$\zeta(0)$ value resulting from the boundary conditions
(1.6)--(1.9). This yields the 1-loop divergence of the
quantum amplitude, and coincides with the conformaly anomaly
in our model. Concluding remarks are presented in section 3.
\vskip 0.3cm
\leftline {\bf 2. $\zeta(0)$ value with non-local boundary conditions}
\vskip 0.3cm
\noindent
As shown in [1, 2, 4, 5], the modes occurring in the expansions
(1.2) and (1.3) obey a coupled set of equations, i.e.
$$
\left({d\over d\tau}-{\Bigr(n+{3\over 2}\Bigr)\over \tau}
\right)x_{np}=E_{np} \; {\widetilde x}_{np}
\eqno (2.1)
$$
$$
\left(-{d\over d\tau}-{\Bigr(n+{3\over 2}\Bigr)\over \tau}
\right){\widetilde x}_{np}=E_{np} \; x_{np}
\eqno (2.2)
$$
where $x_{np}$ denotes $m_{np}$ or $r_{np}$, and
${\widetilde x}_{np}$ denotes ${\widetilde m}_{np}$ or
${\widetilde r}_{np}$. Setting $E_{np}=M$ for simplicity
of notation one thus finds, for all $n \geq 0$, the
solutions of (2.1) and (2.2) in the form
$$
m_{np}(\tau)=\beta_{1,n}\sqrt{\tau}I_{n+1}(M\tau)
+\beta_{2,n}\sqrt{\tau}K_{n+1}(M\tau)
\eqno (2.3)
$$
$$
r_{np}(\tau)=\beta_{1,n}\sqrt{\tau}I_{n+1}(M\tau)
+\beta_{2,n}\sqrt{\tau}K_{n+1}(M\tau)
\eqno (2.4)
$$
$$
{\widetilde m}_{np}(\tau)=\beta_{1,n}\sqrt{\tau}I_{n+2}(M\tau)
-\beta_{2,n}\sqrt{\tau}K_{n+2}(M\tau)
\eqno (2.5)
$$
$$
{\widetilde r}_{np}(\tau)=\beta_{1,n}\sqrt{\tau}I_{n+2}(M\tau)
-\beta_{2,n}\sqrt{\tau}K_{n+2}(M\tau)
\eqno (2.6)
$$
where $\beta_{1,n}$ and $\beta_{2,n}$ are some constants.
The insertion of (2.3)--(2.6) into the boundary conditions (1.6)--(1.9)
leads to the equations (hereafter $b$ and $a$ are the
radii of the two concentric 3-sphere boundaries, with $b>a$,
and we define $\beta_{n} \equiv \beta_{2,n}/\beta_{1,n}$)
$$
I_{n+1}(Mb)+\beta_{n}K_{n+1}(Mb)=0
\eqno (2.7)
$$
for $m_{np}$ and $r_{np}$ modes, and 
$$
I_{n+2}(Ma)-\beta_{n}K_{n+2}(Ma)=0
\eqno (2.8) 
$$
for ${\widetilde m}_{np}$ and ${\widetilde r}_{np}$ modes,
with the same value of $M$ [1]. One thus finds two equivalent
formulae for $\beta_{n}$:
$$
\beta_{n}=-{I_{n+1}(Mb)\over K_{n+1}(Mb)}
={I_{n+2}(Ma)\over K_{n+2}(Ma)}
\eqno (2.9)
$$
which lead to the eigenvalue condition
$$
I_{n+1}(Mb)K_{n+2}(Ma)+I_{n+2}(Ma)K_{n+1}(Mb)=0 .
\eqno (2.10)
$$
The full degeneracy is $2(n+1)(n+2)$, for all $n \geq 0$,
since each set of modes contributes to (2.7) and (2.8) with
degeneracy $(n+1)(n+2)$ [1].

We can now apply $\zeta$-function regularization to evaluate the
resulting conformal anomaly [6], following the algorithm developed 
in [7, 8] and applied several times in the recent literature [9--16].
The basic properties are as follows. Let us denote by $f_{n}$ the
function occurring in the equation obeyed by the eigenvalues by 
virtue of boundary conditions, after taking out fake roots (e.g.
$x=0$ is a fake root of order $\nu$ of the Bessel function
$I_{\nu}(x)$). Let $d(n)$ be the degeneracy of the eigenvalues
parametrized by the integer $n$. One can then define the function
$$
I(M^{2},s) \equiv \sum_{n=n_{0}}^{\infty}
d(n)n^{-2s}\log f_{n}(M^{2})
\eqno (2.11)
$$
and the work in [7, 8] shows that such a function admits an
analytic continuation to the complex-$s$ plane as a meromorphic
function with a simple pole at $s=0$, in the form
$$
``I(M^{2},s)"={I_{\rm pole}(M^{2})\over s}+I^{R}(M^{2})
+{\rm O}(s) .
\eqno (2.12)
$$
The function $I_{\rm pole}(M^{2})$ is the residue at $s=0$,
and makes it possible to obtain the $\zeta(0)$ value as
$$
\zeta(0)=I_{\rm log}+I_{\rm pole}(M^{2}=\infty)
-I_{\rm pole}(M^{2}=0)
\eqno (2.13)
$$
where $I_{\rm log}$ is the coefficient of the $\log(M)$ term
in $I^{R}$ as $M \rightarrow \infty$. The contributions
$I_{\rm log}$ and $I_{\rm pole}(\infty)$ are obtained from the
uniform asymptotic expansions of basis functions as 
$M \rightarrow \infty$ and their order $n \rightarrow \infty$,
whilst $I_{\rm pole}(0)$ is obtained by taking the
$M \rightarrow 0$ limit of the eigenvalue condition, and then
studying the asymptotics as $n \rightarrow \infty$. 
More precisely, $I_{\rm pole}(\infty)$ coincides with the
coefficient of ${1\over n}$ in the expansion as 
$n \rightarrow \infty$ of
$$
{1\over 2}d(n)\log \Bigr[\rho_{\infty}(n)\Bigr]
$$
where $\rho_{\infty}(n)$ is the $n$-dependent term in the
eigenvalue condition as $M \rightarrow \infty$ and
$n \rightarrow \infty$. The $I_{\rm pole}(0)$ value is
instead obtained as the coefficient of ${1\over n}$ in
the expansion as $n \rightarrow \infty$ of
$$
{1\over 2}d(n) \log \Bigr[\rho_{0}(n)\Bigr]
$$
where $\rho_{0}(n)$ is the $n$-dependent term in the
eigenvalue condition as $M \rightarrow 0$ and
$n \rightarrow \infty$ [7, 8, 14].

In our problem, using the limiting form of Bessel functions 
when the argument tends to zero [17], one finds that the
left-hand side of (2.10) is proportional to $M^{-1}$ as
$M \rightarrow 0$. Hence one has to multiply by $M$ to get
rid of fake roots. Moreover, in the uniform asymptotic
expansion of Bessel functions as $M \rightarrow \infty$ and
$n \rightarrow \infty$, both $I$ and $K$ functions contribute
a ${1\over \sqrt{M}}$ factor. These properties imply that
$I_{\rm log}$ vanishes:
$$
I_{\rm log}={1\over 2}\sum_{l=1}^{\infty}2l(l+1)
\Bigr(1-{1\over 2}-{1\over 2}\Bigr)=0 .
\eqno (2.14)
$$
Moreover, 
$$
I_{\rm pole}(\infty)=0
\eqno (2.15)
$$
since there is no $n$-dependent coefficient in 
the uniform asymptotic expansion of (2.10) [7--16].
Last, one finds
$$
I_{\rm pole}(0)=0
\eqno (2.16)
$$
since the limiting form of (2.10) as $M \rightarrow 0$ and
$n \rightarrow \infty$ is
$$
{2\over Ma}(b/a)^{n+1} .
$$
The results (2.14)--(2.16), jointly with the general formula
(2.13), lead to a vanishing value of the 1-loop divergence:
$$
\zeta(0)=0.
\eqno (2.17)
$$
\vskip 0.3cm
\leftline {\bf 3. Concluding remarks}
\vskip 0.3cm
\noindent
To our knowledge, the analysis leading to (2.17) in the
spectral case, had not been performed in the current 
literature. Our detailed calculation shows that, 
in flat Euclidean 4-space, the conformal 
anomaly for a massless spin-1/2 field subject to non-local
boundary conditions of the spectral type on two concentric
3-spheres vanishes, as in the case when the same
field is subject to the local boundary conditions
$$
\sqrt{2} \; {_{e}n_{A}^{\; \; A'}} \; \psi^{A}
= \pm {\widetilde \psi}^{A'} 
\; \; \; \; {\rm on} \; \Sigma_{I} \; {\rm and}
\; \Sigma_{F} \; .
\eqno (3.1)
$$
If (3.1) holds and the spin-1/2 field is massless, the work
in [10] shows in fact that $\zeta(0)=0$.

Backgrounds given by flat Euclidean 4-space bounded by two
concentric 3-spheres are not the ones occurring in the 
Hartle-Hawking proposal for quantum cosmology, where the
initial 3-surface $\Sigma_{I}$ shrinks to a point [18].
Nevertheless, they are relevant for the quantization 
programme of gauge fields and gravitation in the presence
of boundaries [11, 12]. In particular, similar techniques
have been used in section 5 of [16] to study a two-boundary
problem for simple supergravity subject to spectral boundary
conditions in the axial gauge. One then finds the eigenvalue
condition
$$
I_{n+2}(Mb)K_{n+3}(Ma)+I_{n+3}(Ma)K_{n+2}(Mb)=0
\eqno (3.2)
$$
for all $n \geq 0$. The analysis of (3.2) along the same
lines of section 2 shows that transverse-traceless gravitino
modes yield a vanishing contribution to $\zeta(0)$, unlike
transverse-traceless modes for gravitons, which instead
contribute $-5$ to $\zeta(0)$ [12, 16]. 

Thus, the results in [16] seem to
show that, at least in finite regions bounded by one 3-sphere
or two concentric 3-spheres, simple supergravity is not
one-loop finite in the presence of boundaries. Of course,
more work is in order to check this property, and then
compare it with the finiteness of scattering problems suggested
in [19]. Further progress is thus likely to occur by virtue of
the fertile interplay of geometric and analytic techniques [20--24]
in the investigation of heat-kernel asymptotics and (1-loop)
quantum cosmology.
\vskip 0.3cm
\leftline {\bf Acknowledgments}
\vskip 0.3cm
\noindent
I am much indebted to Alexander Kamenshchik for teaching
me all I know about the algorithm used in this paper. 
Joint work with him and with Giuseppe Pollifrone has been
of great help for me.
\vskip 0.3cm
\leftline {\bf References}
\vskip 0.3cm
\item {[1]}
D'Eath P D and Esposito G 1991 {\it Phys. Rev.} 
D {\bf 44} 1713
\item {[2]}
D'Eath P D and Esposito G 1991 {\it Phys. Rev.} D {\bf 43} 3234
\item {[3]}
Atiyah M F, Patodi V K and Singer I M 1975 {\it Math. Proc.
Camb. Philos. Soc.} {\bf 77} 43
\item {[4]}
D'Eath P D and Halliwell J J 1987 
{\it Phys. Rev.} D {\bf 35} 1100
\item {[5]}
Esposito G 1994 {\it Quantum Gravity, Quantum Cosmology and
Lorentzian Geometries} ({\it Lecture Notes in Physics m12})
(Berlin: Springer)
\item {[6]}
Hawking S W 1977 {\it Commun. Math. Phys.} {\bf 55} 133
\item {[7]}
Barvinsky A O, Kamenshchik A Yu and Karmazin I P 1992
{\it Ann. Phys.,} {\it NY} {\bf 219} 201
\item {[8]}
Kamenshchik A Yu and Mishakov I V 1992 {\it Int. J. Mod.
Phys.} A {\bf 7} 3713
\item {[9]}
Kamenshchik A Yu and Mishakov I V 1993 {\it Phys. Rev.}
D {\bf 47} 1380
\item {[10]}
Kamenshchik A Yu and Mishakov I V 1994 {\it Phys. Rev.}
D {\bf 49} 816
\item {[11]}
Esposito G, Kamenshchik A Yu, Mishakov I V and Pollifrone G
1994 {\it Class. Quantum Grav.} {\bf 11} 2939
\item {[12]}
Esposito G, Kamenshchik A Yu, Mishakov I V and Pollifrone G
1994 {\it Phys. Rev.} D {\bf 50} 6329
\item {[13]}
Esposito G, Kamenshchik A Yu, Mishakov I V and Pollifrone G
1995 {\it Phys. Rev.} D {\bf 52} 2183
\item {[14]}
Esposito G, Kamenshchik A Yu, Mishakov I V and Pollifrone G
1995 {\it Phys. Rev.} D {\bf 52} 3457
\item {[15]}
Esposito G and Kamenshchik A Yu 1995 {\it Class. Quantum Grav.}
{\bf 12} 2715
\item {[16]}
Esposito G and Kamenshchik A Yu 1996 One-loop divergences in
simple supergravity: boundary effects {\it Preprint} 
HEP-TH 9604182 (to appear in {\it Phys. Rev.} D)
\item {[17]}
Abramowitz M and Stegun I A 1964 {\it Handbook of Mathematical
Functions with Formulas, Graphs and Mathematical Tables}
(New York: Dover)
\item {[18]}
Hartle J B and Hawking S W 1983 {\it Phys. Rev.} D {\bf 28}
2960
\item {[19]}
D'Eath P D 1996 {\it Supersymmetric Quantum Cosmology}
(Cambridge: Cambridge University Press)
\item {[20]}
Dowker J S 1996 {\it Phys. Lett.} {\bf 366B} 89
\item {[21]}
Kirsten K and Cognola G 1996 {\it Class. Quantum Grav.}
{\bf 13} 633
\item {[22]}
Moss I G and Poletti S J 1994 {\it Phys. Lett.} {\bf 333B} 326
\item {[23]}
Vassilevich D V 1995 {\it J. Math. Phys.} {\bf 36} 3174
\item {[24]}
Gilkey P B 1995 {\it Invariance Theory, the Heat Equation and
the Atiyah-Singer Index Theorem} (Boca Raton, FL: Chemical
Rubber Company)

\bye